\documentclass{ifacconf}

\usepackage{natbib}        

\usepackage{graphics} 
\usepackage{epsfig} 
\usepackage{mathptmx} 
\usepackage{times} 
\usepackage{amsmath} 
\usepackage{amssymb}  
\usepackage{booktabs}
\usepackage{caption}
\usepackage{subcaption}
\usepackage{xcolor}

\usepackage{multirow}
\usepackage{array}

\begin{document}
\begin{frontmatter}

\title{A four-bodies motorcycle dynamic model for observer design\thanksref{footnoteinfo}} 

\thanks[footnoteinfo]{This work was supported by Autoliv and ANRT.}

\author[First,Second]{Tychique K. Nzalalemba} 
\author[Second]{Ziad Alkhoury} 
\author[Third]{Jawwad Ahmed}
\author[Fourth,Fifth]{Mihaly Petreczky}
\author[Fourth,Fifth]{Laurentiu Hetel}
\author[First,Fifth]{Lotfi Belkoura}

\address[First]{University of Lille, 59 000 Lille, France (e-mail: tychique.nzalalembakabwangala.etu@univ-lille.fr).}
\address[Second]{Autoliv Electronics, 95 800 Cergy, France}
\address[Third]{Autoliv Research, 447 37 Vargada, Sweden}
\address[Fourth]{Centrale Lille, 59 651 Villeneuve-d'Ascq, France}
\address[Fifth]{CNRS, UMR 9189 CRIStAL, 59 655 Villeneuve-d'Ascq, France}

\begin{abstract}                
	Motivated by the need to predict dangerous scenarios, this article introduces a non-linear dynamic model for motorcycles consisting of four rigid bodies. Using Jourdain's principle, the model incorporates both longitudinal and lateral dynamics, targeting a balance between numerical complexity and accuracy of representation. The paper further employs the model to design a Luenberger observer based on linear quadratic regulator theory, for estimating physical states  based on sensor measurements. In turn, the state estimates are useful for predicting dangerous scenarios (lowside, highside, fall). The relevance of the approach is demonstrated through simulations of various rectilinear trajectories and a lane-changing scenario using BikeSim simulator.
\end{abstract}


\end{frontmatter}

\section{Introduction}
Safety measures for powered two-wheelers (PTW) represent an important societal problem. In France, despite PTW road users constituting less than 2\% of the motorized traffic, they represented 22\% of fatalities in 2022, [\cite{ONISR2022}]. 
For this reason, improving motorcycle safety is important. One challenge is to detect dangerous situations, using either PTW analytical models or their dynamics' estimations in order to trigger active or passive safety measures. To this end, analytical models of PTW are required in order to predict future safety events, and to be used in estimation algorithms (observers, extended Kalman-filters) for estimating those physical quantities which cannot be directly measured but which are relevant for predicting future danger (fall, etc.).

In turn, due to their unstable and nonlinear behavior, motorcycle modeling is more challenging than for four-wheeled vehicles. From the point of view of mechanics, the main analytical modeling approaches are based either on the Lagrange formalism [\cite{Sharp1971, Sharp2004, Cossalter2002}], or on the Jourdain's principle, [\cite{Nehaoua2013}]. With respect to their use, two types of models can be found in the literature. The first type concerns realistic highly dimensional offline models, used for computing complex dynamics (see [\cite{Sharp2004, Nehaoua2013, Cossalter2006}] and simulators such as BikeSim, MotorcycleMaker, FastBike, etc.). The second type of models are less precise but simpler, which, due to their simplicity, are more suitable for analysis, control or state estimation  [\cite{Sharp1971, Cossalter2002, Corno2012, Bonci2016}]. In particular, these simpler models can be used to design state-estimation and control algorithms which could be implemented real-time.
In this paper, we focus on a model belonging to this second category. More precisely, we focus on models that can be used in state estimation algorithms and thus for monitoring the system dynamics. The interest of such models lies in the detection of potentially dangerous situations.

In the literature, several works have been conducted to estimate only the longitudinal dynamics, [\cite{Panzani2012,Dabladji2015}], or only the lateral dynamics [\cite{Teerhuis2010,Nehaoua2014,Nehaoua2013acc,Ichalal2013,Damon2016,Chenane2012}].
However, results concerning simultaneous estimations of both longitudinal and lateral dynamics are rare [\cite{Dabladji2015icats,Fouka2019icnsc,Caiaffa2023}].


\textbf{Contribution.}
In this paper, we propose a model that describes both lateral and longitudinal dynamics and their interaction, and allows to derive observers using classical linearization and Luenberger observer design. The model is derived by viewing the motorcycle as a rigid body with four components and applying to it the Jourdain's principle. The model presents good compromise between complexity and precision, which is demonstrated by its use for designing Luenberger observer for state-estimation. 

\textbf{Novelty. }
From the point of view of modeling, the main novelty of the proposed model lies in the unique combination of precision and simplicity. In particular, in contrast to this paper, the models in [\cite{Dabladji2015icats,Fouka2019icnsc}] did not capture the interaction between lateral and longitudinal dynamics, but rather modeled these two aspects by two separate models. 
The models of [\cite{Nehaoua2013acc,Ichalal2013,Damon2016,Chenane2012}] and [\cite{Panzani2012,Dabladji2015}] capture only the lateral or longitudinal dynamics respectively, but not the interaction between them. The models used in [\cite{Teerhuis2010, Caiaffa2023}] for Kalman filter do not take into account the dynamics of tire forces (it assumes them to be in equilibrium).
In contrast, our model considers the dynamics of tire forces and the change of the steering angle under the influence of input driving, braking and steering torques. Moreover, the model of this paper uses Jourdain's principle whereas [\cite{Teerhuis2010, Caiaffa2023}] use the Lagrange formalism. The former is often preferred for vehicle modeling [\cite{Rill1994}], as it avoids the need to define explicitly the Lagrangian and to compute its derivatives. 

Concerning the design of the state observer, the novelty of the model is that it allows the use of a simpler observer design for estimating both lateral and longitudinal dynamics. 
In particular, for estimating lateral dynamics or longitudinal dynamics [\cite{Teerhuis2010,Panzani2012,Caiaffa2023}] use extended Kalman-filters,
[\cite{Ichalal2013,Damon2016,Chenane2012}] use sophisticated techniques based on linear matrix inequalities, [\cite{Nehaoua2014,Nehaoua2013acc,Dabladji2015icats}] use sliding mode unknown input observers.
The obtained observers can be challenging to implement. Indeed, the implementation challenges of extended Kalman-filters and sliding mode observers are well-known.
Moreover, the observer gains of [\cite{Ichalal2013,Damon2016,Chenane2012}] obtained from linear matrix inequalities depend on the current state-estimates which makes the implementation harder.
Finally, the conditions of [\cite{Ichalal2013,Damon2016,Chenane2012,Nehaoua2014,Nehaoua2013acc}] might be challenging to verify for all scenarios of interest. 

For estimating both lateral and longitudinal dynamics, [\cite{Fouka2019icnsc,Dabladji2015}] used advanced switched or linear parameter varying observer design techniques, application of which required solving linear matrix inequalities and the satisfaction of rather restrictive conditions. Moreover, the resulted observer is a combination of two parallel observers with observer gains which depend on the state estimates. Such observers are not as easy to implement, and the conditions required for existence of an observer might be challenging to verify.

The paper is structured as follows: in Section \ref{Sec: BasicNotions}, the system description and the paper objectives will be discussed. Section \ref{Sec: Dynamic model} will describe the procedure to derive the four-bodies non-linear motorcycle dynamic model. Section \ref{Sec: ObserverDesign} will use the introduced model to implement an observer using inertial measurement unit data. Finally, results discussion and concluding remarks will close this paper.

\noindent \textbf{Notations}. ${O}_{p \times n}$ denotes a zeros matrix of $p$ lines and $n$ columns, and  ${I}_{p}$ an identity matrix of $p$ lines and $p$ columns.

\section{PTW description and objectives}
\label{Sec: BasicNotions}
In this section, we introduce the mechanical description of the motorcycle based on the four rigid bodies model, and the notations that will be used along the paper. The objectives of the problem are also formalized.

\begin{figure}
	\begin{center}
		\includegraphics[width=8.9cm]{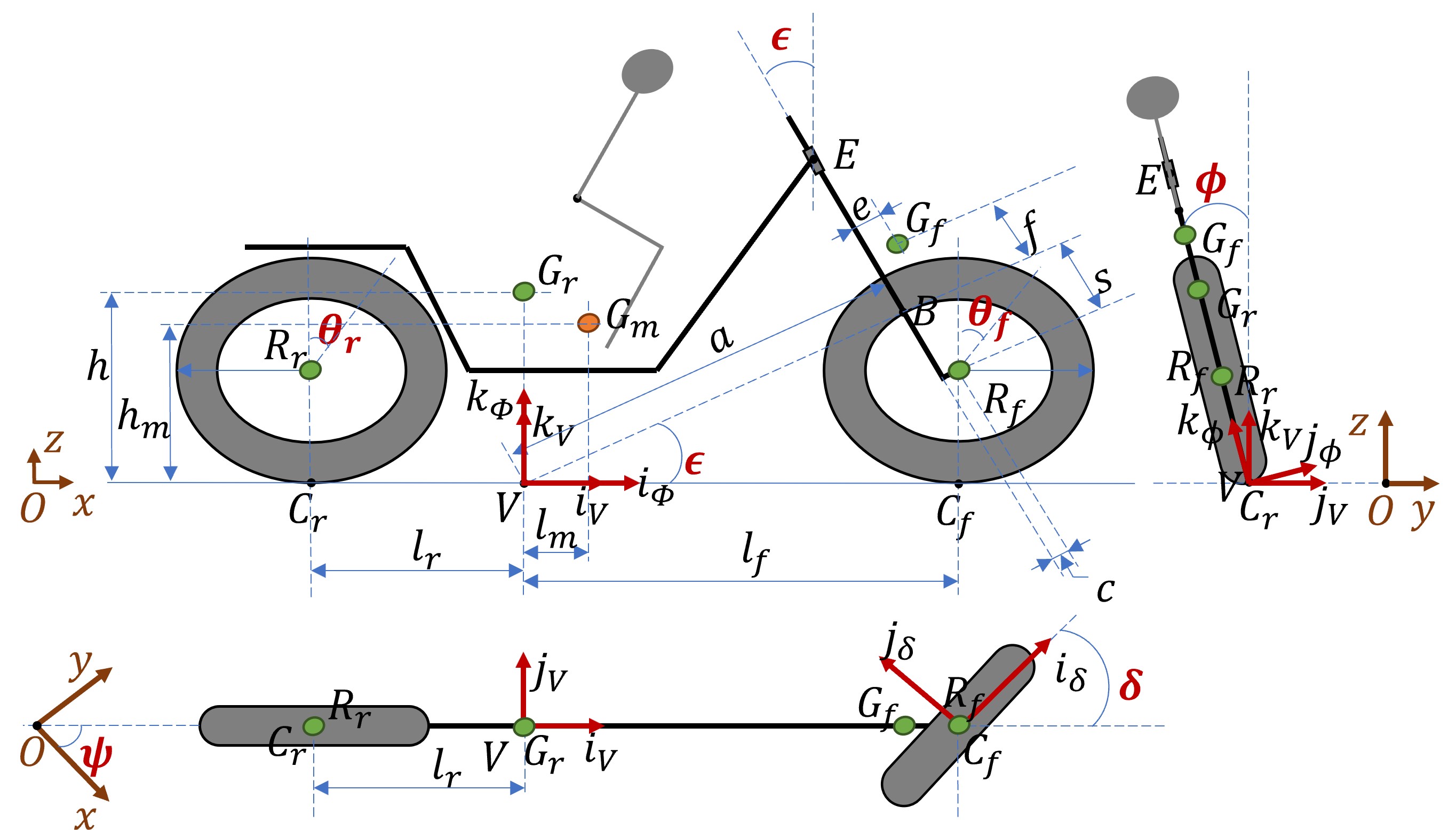}   
		\caption{PTW model description} 
		\label{fig2}
	\end{center}
\end{figure}

\emph{Four-bodies motorcycle description}
The motorcycle (see Fig.\ref{fig2}) is represented as a set of four rigid bodies composed of:
\begin{itemize}
	\item the front body $G_{f}$ which includes the steering assembly, the front suspension and braking system;
	\item the rear body $G_{r}$, incorporating the main frame, the engine, the tank, the saddle, the swinging arm, the rear suspension and braking assemblies, and the rider;
	\item the front wheel $R_{f}$ and the rear wheel $R_{r}$.
	\item $G_{m}$ is the center of mass of the whole motorcycle.
\end{itemize}
We then have seven degrees of freedom, which are: the longitudinal ($x$) and lateral ($y$) displacements, the yaw ($\psi$), roll ($\phi$) and steering ($\delta$) angles, and the front ($\theta_{f}$) and rear ($\theta_{r}$) wheels spinning angles.

The PTW longitudinal and lateral velocities are denoted $v_{x}$ and $v_{y}$, respectively. It is also assumed that the PTW motion is driven by the steering torque $\tau$ applied by the rider, the engine propulsive torque $\tau_{D}$, the front and the rear braking torques, $\tau_{B_f}$ and $\tau_{B_r}$, applied at the front and rear wheels, respectively.

To account for the interactions between the motorcycle and the ground during motion, the tire forces need to be modeled. 
In terms of control theory, the effects of torques are considered as inputs and the tire dynamics are part of the model states dynamics.

\subsubsection{Tire dynamics.}
In the analysis of tire dynamics (as shown in Fig. \ref{fig4}), we consider three forces acting on the contact between the tire and the ground [\cite{Pacejka2006}]:
\begin{figure}
	\begin{center}
		\includegraphics[width=5cm]{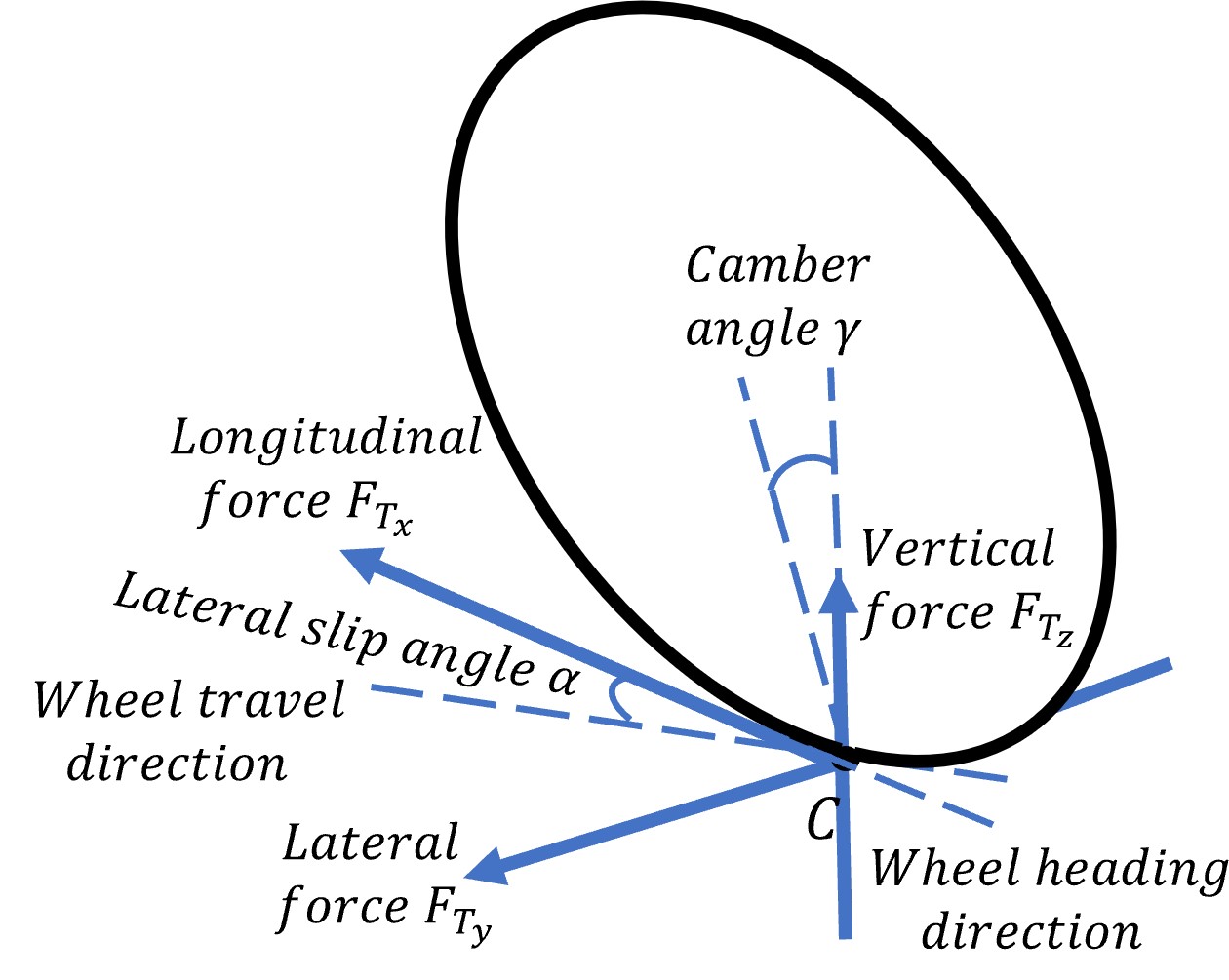}   
		\caption{Motorcycle and tire forces description} 
		\label{fig4}
	\end{center}
\end{figure}

\subsubsection{Vertical tire forces.}
Normal loads ($F_{T_{f_{z}}}$ and $F_{T_{r_{z}}}$) consist of two components: \textit{static load} which is associated with the motorcycle's weight, including the rider, passenger, and luggage, and \textit{dynamic load} that depends on factors such as road inclination, aerodynamic drag, etc.
For simplicity, we assume that the vertical tire forces are made of the static load only.

\subsubsection{Longitudinal and lateral tire forces.}
These forces are due to presence of the driving and braking torques causing forward velocity, accelerations or decelerations. In particular, forward motions and longitudinal slips generate longitudinal tire forces $F_{T_{f_{x}}}$ and $F_{T_{r_{x}}}$, and cornering, side slips and camber angles induce lateral tire forces  $F_{T_{f_{y}}}$ and $F_{T_{r_{y}}}$. The manner to describe them will be discussed in section \ref{Sec: Dynamic model}.

\subsubsection{Objectives.}
This paper objectives are formalized as follows:\\
		\textbf{(1)}
		provide a PTW dynamic model based on the four-bodies description presented in Fig. \ref{fig2}; \\
		\textbf{(2)}  assuming the longitudinal acceleration $\dot{v}_x$, the lateral acceleration $\dot{v}_y$, the roll-rate $\dot{\phi}$ and the yaw-rate $\dot{\psi}$ are measured by an inertial measurement unit sensor, design an observer that allows to estimate the yaw $\psi$, roll $\phi$, steering $\delta$ angles, the longitudinal and lateral velocities $v_{x}$ and $v_{y}$, the steering rate $\dot{\delta}$, the wheels' spinning rates $\dot{\theta}_{f}$ and $\dot{\theta}_{r}$, and the tire forces ${F}_{T_{f_x}}$, ${F}_{T_{r_x}}$, ${F}_{T_{f_y}}$, ${F}_{T_{r_y}}$.

\section{Dynamic model derivation}
\label{Sec: Dynamic model}
The PTW model introduced in this paper is based on Jourdain's principle [\cite{Rill1994, Nehaoua2013}]. The goal is to provide the main steps that allow to derive a motorcycle dynamical model of the form ${\cal M} \dot{v} = Q$,
where $v = \frac{d q}{dt}$ and  $\dot{v}=\frac{d v}{dt}$;
$q$ is the \emph{generalized coordinates vector}, $v$ \emph{the generalized velocity vector}, $\dot{v}$ is \emph{generalized acceleration}, $Q$ the generalized effort vector and ${\cal M}$ the generalized mass matrix [\cite{Rill1994, Roberson1988}]. For the case under study, $q$ is given by $q = $[$x$, $y$, $\psi$, $\phi$, $\delta$, $\theta_{f}$, $\theta_{r}$]$^{T}$ and $v$ by $v = $[$v_{x}$, $v_{y}$, $\dot{\psi}$, $\dot{\phi}$, $\dot{\delta}$, $\dot{\theta}_{f}$, $\dot{\theta}_{r}$]$^{T}$.
Notice that  $\frac{d}{dt}{q}(t) = v$, i.e., the generalized velocity is the time derivative of the generalized coordinated. 
We denote by $\dot{v}=\frac{d v}{dt}$ and we refer to it as \emph{generalized acceleration.}
Adding to this a classical model representing tire kinematics, [\cite{Pacejka2006}], a non-linear model is deduced. This latter can be used for observer design.

\subsubsection{Modeling assumptions.}
The following assumptions are used while deriving the motorcycle four-bodies non-linear model:
\begin{itemize}
	\item[$\blacksquare$] \pmb{$A_{1}$}: The four-bodies are symmetrical in the longitudinal motion plane associated to each body.
	\item[$\blacksquare$] \pmb{$A_{2}$}: The road is flat.
	\item[$\blacksquare$] \pmb{$A_{3}$}: The aerodynamic lift force, the tire moments, the rolling resistance forces, the pitch and the suspensions dynamics are not taken into account.
	\item[$\blacksquare$] \pmb{$A_{4}$}: Only the static loads are considered in the vertical tire forces computation.
\end{itemize}

\subsubsection{Preliminaries on Jourdain's principle.}
The motorcycle is viewed as a collection of four rigid bodies, where the body $i$ has a mass $m_{i}$, an inertia moment ${\cal J}_{i}$ and its center of mass at the point $i$, for  $i \in \{G_{f},G_{r},R_{f},R_{r}\}$. That is, each body is identified by its center of mass.

Let ${\mathcal R}_{O} (O,x,y,z)$ be an inertial reference frame.
In addition, we consider  ${\mathcal R}_{V} (V,i_{V},j_{V},k_{V})$, centered at the point $V$, which moves with the motorcycle. The point $V$ is chosen as the intersection between the PTW longitudinal plane of symmetry, the horizontal ground plane and the vertical axis passing through the rear body center of mass $G_r$. In addition, we consider the local frames  ${\mathcal R}_{i} (i,i_{i},j_{i},k_{i})$ of the body $i$ centered at its center of mass, for $i \in \{G_{f},G_{r},R_{f},R_{r}\}$. The orientation of these frames is the same as described in [\cite{Damon2018}], and it is illustrated on Fig. \ref{fig2}. 

We denote by $r_{Xi}^{Y}$  the position of the center of mass of the body $i$ in the frame ${\mathcal R}_{Y}$ relative to the point $X$, and by ${\Re_{ZY}}$ the rotation matrix transforming coordinates in frame $\mathcal{R}_Y$ to frame $\mathcal{R}_Z$.
Note that $r_{Xi}^Y$ and ${\Re_{ZY}}$ depend on time $t$ via the generalized coordinate $q$, i.e. $r_{Xi}^Y$ and ${\Re_{ZY}}$ are functions of $q$, and $q$ is function of $t$. In the sequel, we will view both $r_{Xi}^Y$ and ${\Re_{ZY}}$ as functions in an independent variable $q$ and functions of time, by viewing $q$ as a time function. In particular, we consider derivatives of $r_{Xi}^Y$  and ${\Re_{ZY}}$ with respect to the variable $q$ and with respect to the time. 

Furthermore, we denote by $v_{Xi}^Y(q,v)$ and $\omega_{XZ}(q,v)$ the linear and angular velocities induced by the position vector $r_{Xi}^Y$ and the rotation $\Re_{ZY}$ respectively, i.e.,
\begin{equation} \label{Eq01}
	\begin{split}
		&  v_{Xi}^{Y}(q,v)=
		{\displaystyle \frac{d}{d t}r_{Xi}^{Y}(q)}, ~  
		\omega_{ZY}(q,v) = \begin{bmatrix} \omega_1 & \omega_2 & \omega_3 \end{bmatrix}^T, \\
		& \frac{d {\Re}_{ZY}(q)}{dt}{\Re}_{ZY}^T(q)=\begin{bmatrix} 0 & -\omega_3  &  \omega_2 \\ \omega_3 & 0 & -\omega_1 \\ -\omega_2 & \omega_1  & 0\ \end{bmatrix}.
	\end{split}
\end{equation}
 Note that explicit expressions of $v_{Xi}^Y(q,v)$ and $\omega_{XZ}(q,v)$ in terms of $q$ and $v$
 can readily be computed from the explicit expressions for
$r_{Xi}^{Y}(q)$ and ${\Re}_{ZY}(q)$ as function of $q$, using the chain rule and the definition of $v$ and $q$.
 In the sequel, we view  $v_{Xi}^Y(q,v)$ and $\omega_{XZ}(q,v)$ as functions of $q$ and $v$. 

	It is well known (see [\cite{Arnold1978}]) that any point of the body $i$ has the same angular velocity $\omega_i(q,v)$ in the local frame $\mathcal{R}_i$,
	which depends on the generalized coordinate $q$ and velocity $v$.
	Thus, any point of the rigid body has the same  angular velocity $\omega_{Oi}^O(q,v)$ in the referential frame, referred to as the \emph{angular velocity of the rigid body $i$}, which is defined as follows:
       \begin{equation} \label{Eq-01}
	\begin{split}
		\omega_{Oi}^{O}(q,v)= {\Re}_{Oi} \omega_{i}(q,v)+\omega_{Oi}(q,v),
	\end{split}
 \end{equation}
 i.e., it is the sum of the rotated version of the common angular velocity in the local frame and the angular velocity induced by the rotation from the local to the inertial 
 reference frame.

 Jourdain's principle establishes a differential equations for the angular  velocities $\omega_{Oi}^{O}(q,v)$ for each body $i$ and	the linear velocity $v_{Oi}^{O}(q,v)$ of the center of mass of the body $i$. 
 More precisely, following [\cite{Arnold1978}], we define the \textit{linear acceleration}  and \textit{angular acceleration} of the body $i$ as:
 \begin{equation} \label{Eq02}
       a_{Oi}^{O}(q,v,\dot{v}) = {\displaystyle \frac{d v_{Oi}^{O}(q,v)}{d t}}, ~ \gamma_{Oi}^{O}(q,v,\dot{v})=\frac{d \omega_{Oi}^O(q,v)}{dt}.
 \end{equation}
By [\cite{Rill1994}], for suitable functions $a_{Ri}^{O}(q,v)$ (called \emph{residual acceleration})  and $\gamma_{Ri}^{O}(q, v)$  (called \emph{residual angular acceleration}), it follows that:
\begin{equation} \label{Eq03}
	\begin{split}
		& a_{Oi}^{O}(q,v,\dot{v}) = a_{Ri}^{O}(q, v) - {\displaystyle \frac{d v_{Oi}^{O}}{d v}}(q, v)\dot{v}, \\
		& \gamma_{Oi}^{O}(q,v,\dot{v}) = \gamma_{Ri}^{O}(q, v) - {\displaystyle \frac{d \omega_{Oi}^{O}}{dv}}(q, v) \dot{v}. 
	\end{split}
\end{equation}

Let $F_{i}$ be the sum of the  external forces acting on the rigid body $i$ and let  $M_{i}(t)$ be the sum of the corresponding moments. 
The \emph{Jourdain's principle} states that:
\begin{equation} \label{Eq05}
	{\cal M} {\displaystyle \frac{d}{d t}} v(t) = Q_a - Q_r,
\end{equation}
where
		the generalized mass matrix ${\cal M}$ is  
		\begin{equation} \begin{split} {\cal M} &= \sum_{i=1}^{4} \left\{ m_i \left( \frac{d v_{Oi}^{O}}{d v} \right)^T \frac{d v_{Oi}^{O}}{d v}  +  \left( \frac{d \omega_{Oi}^{O}}{d v} \right)^T \tilde{\cal J}_i \frac{d \omega_{Oi}^{O}}{d v} \right\} \end{split}, \tag{5a} \label{Eq5a}  \end{equation}
		the generalized vector of external efforts $Q_{a}$ is defined by
	\begin{equation}
		Q_{a} = \sum_{i=1}^{4}\left\{\left( \frac{d v_{Oi}^{O}}{d v} \right)^{T} F_{i} + \left( \frac{d \omega_{Oi}^{O}}{d v} \right)^{T} M_{i} \right\},
		\tag{5b} \label{Eq5b}
	\end{equation}
	the generalized vector of residual efforts $Q_{r}$ is defined as
	{\small
	\begin{equation}
		Q_{r} = \sum_{i=1}^{4} \left\{ \left( m_i  \frac{d v_{Oi}^{O}}{d v} \right)^T a_{Ri}^{O}  + 
		\left( \frac{d \omega_{Oi}^{O}}{d v} \right)^T   \left ( \tilde{\cal J}_i \gamma_{Ri}^{O} + 
		\omega_{Oi}^{O} \times \tilde{\cal J}_i \omega_{Oi}^{O} \right) \right\},
		\tag{5c} \label{Eq5c} 
	\end{equation}
	}
	and the inertial tensor  $\tilde{\cal J}_i$ is given by $\tilde{\cal J}_i = {\Re}_{Oi} {\cal J}_i {\Re}_{Oi}^{T}$. 

\subsubsection{Explicit equations using Jourdain's principle.}
In order to be able to use (\ref{Eq05}), we present explicit expressions for the rotation $\Re_{Oi}$, the linear and angular velocities 
$v_{Oi}^O$ and $\omega_{Oi}^{O}$ and the forces $F_{i}$ and moments $M{i}$. To this end, it is convenient to use the mobile reference frame $\mathcal{R}_{V}$ and notice that:


\begin{equation} \label{Eq1}
	\begin{split}
		& r_{Oi}^{O} = r_{OV}^{O} + {\Re}_{OV} r_{Vi}^{V}, 
			~ \Re_{Oi}=\Re_{OV} \Re_{Vi} \\
			&			 
		~ r_{OV}^{O} = \begin{bmatrix} x \\ y \\ 0 \end{bmatrix}, 
			~  {\Re}_{OV} = \begin{bmatrix}, ~ 
		\cos(\psi) & -\sin(\psi) & 0 \\ \sin(\psi) & \cos(\psi) & 0 \\ 0 & 0 & 1 
	\end{bmatrix}.
       \end{split}
\end{equation}
Note that since the PTW is assumed to be in permanent contact with the ground, the last coordinate of $r_{OV}$ is equal to $0$.
	It follows that $r_{Oi}^{O}(q)$  and $\Re_{Oi}(q)$ can easily be computed from the knowledge of $r_{Vi}^V(q)$ and 
	$\Re_{Vi}(q)$. Then, $v_{Oi}^{O}(q,v)$ can readily be computed by differentiating $r_{Oi}^{O}(q)$ and the chain rule, and $\omega_{Oi}^{O}(q,v)$ can easily be computed from $\omega_{Vi}(q,v)$, $\Re_{Vi}(q)$ and $\omega_i(q,v)$ and \eqref{Eq-01}. 
	We define the latter as in [\cite{Damon2018}], see table \ref{Tab0}, where

\begin{equation*} 
	\begin{split}
		&		{\Re}_{\phi}^{1} = \begin{bmatrix}
	1 & 0 & 0 \\ 0 & \cos(\phi) & -\sin(\phi) \\ 0 & \sin(\phi) & \cos(\phi)
	\end{bmatrix}, ~
				{\Re}_{\epsilon}^{2} = \begin{bmatrix}
		\cos(\epsilon) & 0 & -\sin(\epsilon) \\ 0 & 1 & 0 \\ \sin(\epsilon) & 0 & \cos(\epsilon)
	\end{bmatrix} \\
	& {\Re}_{\delta}^{3} = \begin{bmatrix}
		\cos(\delta) & -\sin(\delta) & 0 \\ \sin(\delta) & \cos(\delta) & 0 \\ 0 & 0 & 1 
	\end{bmatrix}.
     \end{split} 
\end{equation*}

\begin{table}[hb]
	\captionsetup{width=\textwidth}
	\caption{Rigid bodies' linear and angular velocities} \label{Tab0}
	\begin{tabular}{ | m{0.8cm} | m{3.2cm}| m{1.5cm} | m{1.2cm} | } 
		\hline
		Body $i$ & $r_{Vi}^{V}$ & ${\Re}_{Vi}$ & $\omega_i$ \\
		\hline
		$G_r$ &  ${\Re}_{\phi}^{1} \begin{bmatrix} 0 & 0 & h \end{bmatrix}^T$ &   $R_{\phi}$ &  0 \\
		\hline
		$R_r$ &   ${\Re}_{\phi}^{1} \begin{bmatrix} -l_r & 0 & R_r \end{bmatrix}^{T}$ &   $R_{\phi}$ & $\begin{bmatrix} 0 &  \dot{\theta}_r & 0 \end{bmatrix}^T$  \\
		\hline
		$G_f$ &   ${\Re}_{\phi}^{1} {\Re}_{\epsilon}^{2} \left( \begin{bmatrix} a \\ 0 \\ 0 \end{bmatrix} +{\Re}_{\delta}^{3} \begin{bmatrix} e \\ 0 \\ f \end{bmatrix}
		\right)$  & ${\Re}_{\phi}^{1} {\Re}_{\epsilon}^{2}{\Re}_{\delta}^{3}$ & 0 \\
		\hline
		$R_f$ & ${\Re}_{\phi}^{1} {\Re}_{\epsilon}^{2} \left(\begin{bmatrix} a \\ 0 \\ 0 \end{bmatrix} + {\Re}_{\delta}^{3} \begin{bmatrix} c \\ 0 \\ - s \end{bmatrix}\right)$ &  ${\Re}_{\phi}^{1} {\Re}_{\epsilon}^{2}{\Re}_{\delta}^{3}$  & $\begin{bmatrix} 0 &  \dot{\theta}_f & 0 \end{bmatrix}^T$   \\
		\hline
	\end{tabular}
\end{table}

The forces acting of the rigid body $i$ is composed of the gravitational force $F_{g_i}$ acting on each body $i$, $i \in \{G_{f},G_{r},R_{f},R_{r}\}$, the 
aerodynamic drag force $F_{a_d}$, and the tire forces $F_{T_k}$,  $k \in \{f, r\}$ acting on the rear and front wheel respectively:
\begin{equation*}
\begin{split}
F_{g_i} = \begin{bmatrix} 0 \\ 0 \\ -m_i g \end{bmatrix}, ~
F_{a_d} = \begin{bmatrix} - {\displaystyle \frac{1}{2}} \rho_{air} C_{d} A_{v} (v_x^V)^2 \\ 0 \\ 0 \end{bmatrix}, ~
F_{T_k} = \begin{bmatrix} F_{T_{k_x}} \\ F_{T_{k_y}} \\ F_{T_{k_z}} \end{bmatrix}.\end{split}
\end{equation*}
We assume that the moments of these forces are zero, except that of $F_{T_k}$ which applies to the contact point $C_k$ of the rear and front wheel for $k=r,f$ respectively. 
The moments are the engine and braking torques acting on the rear wheel and the steering torque acting on the front body $G_f$.
The forces corresponding to the latter moments are not affecting the linear motion of the corresponding body, and hence they are not integrated into the Jourdain's formalism. The forces and moments acting on each body are summarized in table \ref{Tab01}.
\begin{table}[h]
	\captionsetup{width=\textwidth}
	\caption{Forces \& moments acting on rigid bodies} \label{Tab01}
	\begin{tabular}{ | m{0.8cm} | m{1.91cm}| m{4.85cm} |} 
		\hline
		Body $i$ & $F_i$ & $M_i$   \\
		\hline
		$G_r$ &  $\Re_{OV}(F_{g_{G_r}}+F_{a_d})$  &  0 \\
		\hline
		$R_r$ &     $\Re_{OV} \left( F_{g_{R_r}} + F_{T_{r}} \right) $  &   {\small ${\Re}_{OV} \left( \begin{bmatrix} 0 \\ \tau_D + {\tau_{B_r}} \\ 0 \end{bmatrix}  + \Re^1_{\phi}  \begin{bmatrix} 0 \\ 0 \\ - R_r \end{bmatrix} \times F_{T_r} \right) $}  \\
		\hline
		$G_f$ &    $\Re_{OV}F_{g_{G_f}}$  &  {\small ${\Re}_{OV} \begin{bmatrix} 0 & 0 & \tau - K_\delta \dot{\delta} \end{bmatrix}^T$ } \\
		\hline
		$R_f$ &  $\Re_{OV}\left( F_{g_{R_f}} + F_{T_{f}} \right)$  &   {\small ${\Re}_{OV} \left(\begin{bmatrix} 0 \\ {\tau_{B_f}} \\  0 \end{bmatrix} + \Re^1_{\phi}\Re^3_{\delta}   \begin{bmatrix} 0 \\ 0 \\ - R_f \end{bmatrix} \times F_{T_f}  \right)$} \\
		\hline
	\end{tabular}
\end{table}

Finally, the tire forces are define as follows. 
Since the dynamic normal loads are not considered, using Assumptions $A_{2} - A_{4}$, the vertical forces are given by:
\begin{equation} \label{Eq10}
	F_{T_{f_{z}}}= m g {\displaystyle \frac{l_r + l_m}{l_r + l_f}} ; \
	F_{T_{r_{z}}}= m g {\displaystyle \frac{l_f - l_m}{l_r + l_f}},
\end{equation}
where $k \in \{f, r\}$ is the subscript for the front and rear tires.

The longitudinal and lateral tire dynamics are based on the classical model presented in [\cite{Cossalter2006}]:
\begin{align} 
	& \dot{F}_{T_{k_{x}}}=F_{x,k}, \quad \dot{F}_{T_{k_{y}}}=F_{y,k}, \label{Eq9} \\
	& F_{x,k}= \frac{v_x^V}{ \sigma_{k_{x}} } \left( F_{T_{k_{x_0}}}-F_{T_{k_{x}}} \right), \quad  F_{y,k}= \frac{v_y^V}{\sigma_{k_{y}}} \left( F_{T_{k_{y_0}}}-F_{T_{k_{y}}} \right), \nonumber 
\end{align}
where 
where $k \in \{f,r\}$, $\sigma_{k_{x}}$ and $\sigma_{k_{y}}$ are the longitudinal and lateral relaxation lengths, $F_{T_{k_{x}}}$ and $F_{T_{k_{y}}}$ are instantaneous tire forces, $F_{T_{k_{x_0}}}$ and $F_{T_{k_{y_0}}}$ are steady states tire forces. The latter follow Pacejka magic formula [\cite{Pacejka2006}]:
\begin{equation} \label{Eq4c}
	y(\mu) = {\cal D} \sin[{\cal C} \arctan({\cal B} \mu - {\cal E} ({\cal B} \mu - \arctan({\cal B} \mu) ) )],
\end{equation}
where $y \in \{ F_{T_{k_{x_0}}}, F_{T_{k_{y_0}}} \}$ is the steady state tire force, $k \in \{f, r\}$, ${\cal D}$ the peak factor, ${\cal C}$ the shape factor, ${\cal B}$ the stiffness factor, ${\cal E}$ the curvature factor and $\mu = \{ \kappa_k, \alpha_k, \gamma_k \}$ the slippage ratios composed of the longitudinal slip, the side slip and the camber angle, respectively. ${\cal D}$, ${\cal C}$, ${\cal B}$ and ${\cal E}$ depends on the normal loads and the slippage ratios.

\subsubsection{Non-linear dynamic model derivation.}
By combining equations (\ref{Eq03}) - (\ref{Eq10}), one can derive a model of the form:
\begin{equation} \label{Eq13b}
	{\cal M}(\psi, \phi, \delta) \dot{v} = Q(v,u),
\end{equation}
where ${\cal M}$ and $Q$  are defined in (\ref{Eq05}) and $u = $[$\tau$, $\tau_{D}$, $\tau_{B_f}$, $\tau_{B_r}$]$^{T}$ is the system input vector.

After combining the expressions (\ref{Eq9}) and (\ref{Eq13b}), one finds a non-linear model driven by:
\begin{equation} \label{Eq3e}
	{\cal M}_{ext}(X_{ext})\dot{X}_{ext} = Q_{ext}(X_{ext},u),
\end{equation}

where  $X_{ext} = \begin{bmatrix} \psi, & \phi, & \delta, & v^T, & {F}_{T_{f_x}}, & {F}_{T_{r_x}}, & {F}_{T_{f_y}}, & {F}_{T_{r_y}} \end{bmatrix}^T$
			is the extended state-space vector of the model;
\[ {\cal M}_{ext}(X_{ext}) = \begin{bmatrix}
			I_{3} & {O}_{3 \times 7} & {O}_{3 \times 4}\\
			{O}_{7 \times 3} & {\cal M}(\psi,\phi,\delta) & {O}_{7 \times 4} \\
			{O}_{4 \times 3} & {O}_{4 \times 7} & {I}_{4} 
\end{bmatrix} \]
is the extended mass matrix; and
\begin{align*}
	Q_{ext}(X_{ext},u) & = \begin{bmatrix}
		\dot{\psi}, & \dot{\phi}, & \dot{\delta}, & Q(v,u)^T,  &  F_{x,f}, & F_{x,r}, &  F_{y,f}, & F_{y,r}
\end{bmatrix}^T 
\end{align*}
is the combination of the generalized effort vector and the right-hand side of the differential equations \eqref{Eq9} describing
	the evolution of tire forces. 

Using the inverse of the generalized mass matrix ${\cal M}_{ext}(X)$, the generic formulation of a non-linear system below is deduced:
\begin{equation} \label{Eq40c}
	\dot{X}_{ext} = h(X_{ext}, u),
\end{equation}
where $h(X_{ext}, u) = {\cal M}_{ext}^{-1}(X_{ext}) {Q}_{ext}(X_{ext},u)$.

\section{OBSERVER DESIGN}
\label{Sec: ObserverDesign}

Let $(X_{ext}^{*}, u^{*})$ be an equilibrium point of (\ref{Eq40c}) along a rectilinear trajectory. Since rectilinear trajectories are considered, $X_{ext}^{*}$ and $u^{*}$ are of the following form: $X_{ext}^{*} = $[0, 0, 0, $v_{x}^{*}$, 0, 0, 0, 0, $\dot{\theta}_{f}^{*}$, $\dot{\theta}_{r}^{*}$, ${F}_{T_{f_x}}^{*}$, ${F}_{T_{r_x}}^{*}$, 0, 0]$^{T}$ and $u^{*} = $[0, $\tau_{D}^{*}$, 0, 0]$^{T}$.

Consider the coordinate transformation $x = X_{ext} - X_{ext}^{*}$. Using the IMU sensor measurements $s$, where $s = $[$\dot{v}_{x}$, $\dot{v}_{y}$, $\dot{\phi}$, $\dot{\psi}$]$^{T}$, one would like to provide an estimate $\hat{x}$ of $x$. Therefore, the following classical observer structure will be considered:
\begin{equation} \label{Eq50a}
	\dot{\hat{x}}(t) = (A - G C) \hat{x}(t) + G s(t) + (B - G D) u(t), \\
\end{equation}
where $A = {\displaystyle \frac{\partial h}{\partial X_{ext}}} \Bigg|_{X_{ext} = X_{ext}^{*}; u = u^{*}}$, 
\vspace{0.2cm}
$B = {\displaystyle \frac{\partial h}{\partial u}} \Bigg|_{X_{ext} = X_{ext}^{*}; u = u^{*}}$, \\ $ C = \begin{bmatrix} H A \\ F \end{bmatrix}$,
\vspace{0.2cm}
$D = \begin{bmatrix} H B \\ O_{2 \times 4} \end{bmatrix}$, $F = \begin{bmatrix} O_{2 \times 5} & I_{2} & O_{2 \times 7} \end{bmatrix}$,\\ $H = \begin{bmatrix} O_{2 \times 3} & I_{2} & O_{2 \times 9} \end{bmatrix}$.

The observer gain $G$ is designed based on the linear quadratic regulator theory, [\cite{Hespanha2018}]. In the observer case, the linear quadratic regulator algorithm aims at minimizing a quadratic cost function, which depends on the state estimation error.


\section{RESULTS AND DISCUSSION}

To illustrate our method, we considered a Suzuki GSX-R1000 motorcycle characterized by the numerical values provided in Table \ref{Tab1}. The observer (\ref{Eq50a}) and the model (\ref{Eq3e}) were implemented for two different scenario types:
\begin{itemize}
	\item Rectilinear motion: this case scenario fits with observer design and motorcycle modeling assumptions;
	\item Overtaking scenario: here, the observer is tested outside of its design scope. In fact, this case implies changing the road lane which induces the lateral dynamics in the PTW. 
\end{itemize}

\begin{table}[hb]
	\captionsetup{width=\textwidth}
	\caption{PTW parameters\label{Tab1}}
		\begin{tabular}{|c|}
			\hline
			All the numerical values are in SI units\\
			\hline
			$l_f = 0.727$, $l_r = 0.643$, $h = 0.5712$, $R_r = 0.297$, $R_f = 0.282$, $s = 0.0381$, \\
			$e = 0.1548$, $f = 0.1893$, $a = 0.7523$, $c = 0.0265$, $\sigma_{f_{x}} = \sigma_{r_{x}} = 0.025$,\\
			$\sigma_{f_{y}} = \sigma_{r_{y}} = 0.200$, $m_r = 257.06$, $m_f = 24.24 $, $m_{R_f} = 7$, $m_{R_r} = 14.7$,\\
			$\epsilon = 24^{o}$, $C_{d} = 0.52$, $A_{v} = 0.6$, $\rho_{air} = 1.206$, $K_{\delta} = 12.6738$,\\
			${\cal J}_{R_r} = \begin{bmatrix} 0 & 0 & 0 \\ 0 & 0.638 & 0 \\ 0 & 0 & 0 \end{bmatrix}$, ${\cal J}_{G_r} = \begin{bmatrix} 19.466 & 0 & -3.659 \\ 0 & 46.293 & 0 \\ -3.659 & 0 & 31.316 \end{bmatrix}$,\\
			${\cal J}_{R_f} = \begin{bmatrix} 0 & 0 & 0 \\ 0 & 0.484 & 0 \\ 0 & 0 & 0 \end{bmatrix}$, \textcolor{black}{ ${\cal J}_{G_f} = \begin{bmatrix} 1.965 & 0 & -0.270 \\ 0 & 2.333 & 0 \\ -0.270 & 0 & 0.537 \end{bmatrix}$}\\
			\hline
		\end{tabular}
\end{table}

\begin{figure}[h]
	\centering
	\begin{subfigure}[b]{0.225\textwidth}
		\centering
		\includegraphics[width=\textwidth]{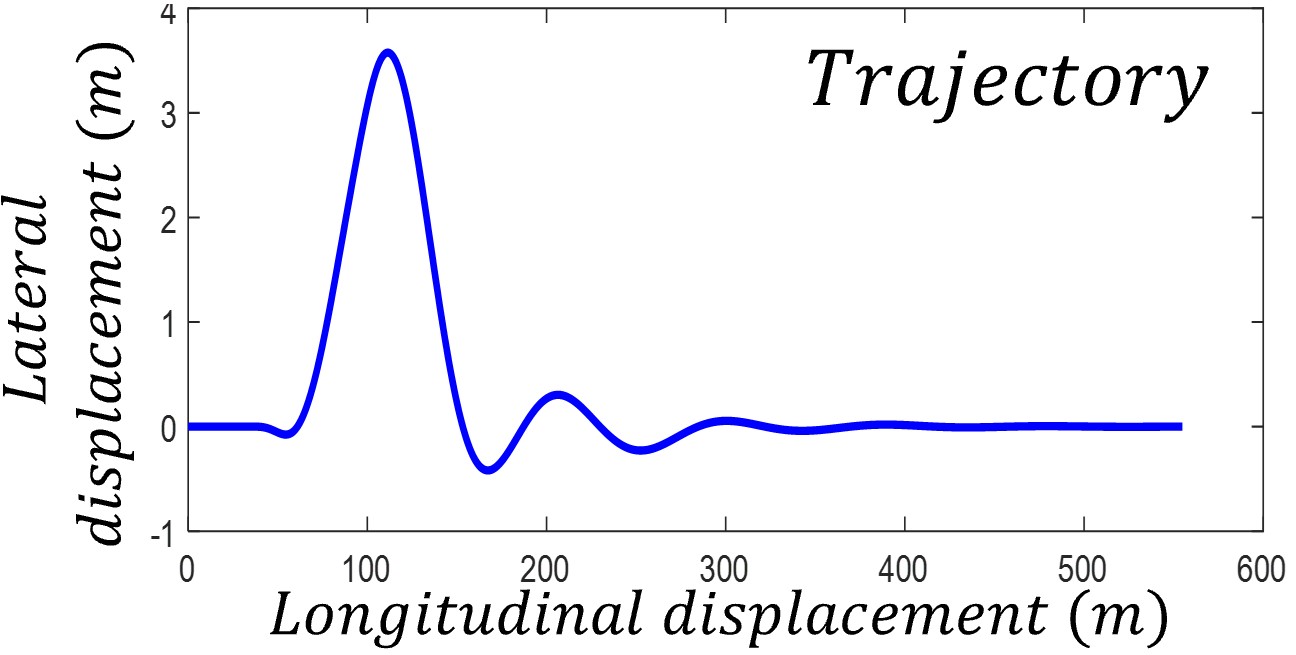}
		\caption{Overtaking trajectory}
		\label{fig12a}
	\end{subfigure}
	\hfill
	\begin{subfigure}[b]{0.235\textwidth}
		\centering
		\includegraphics[width=\textwidth]{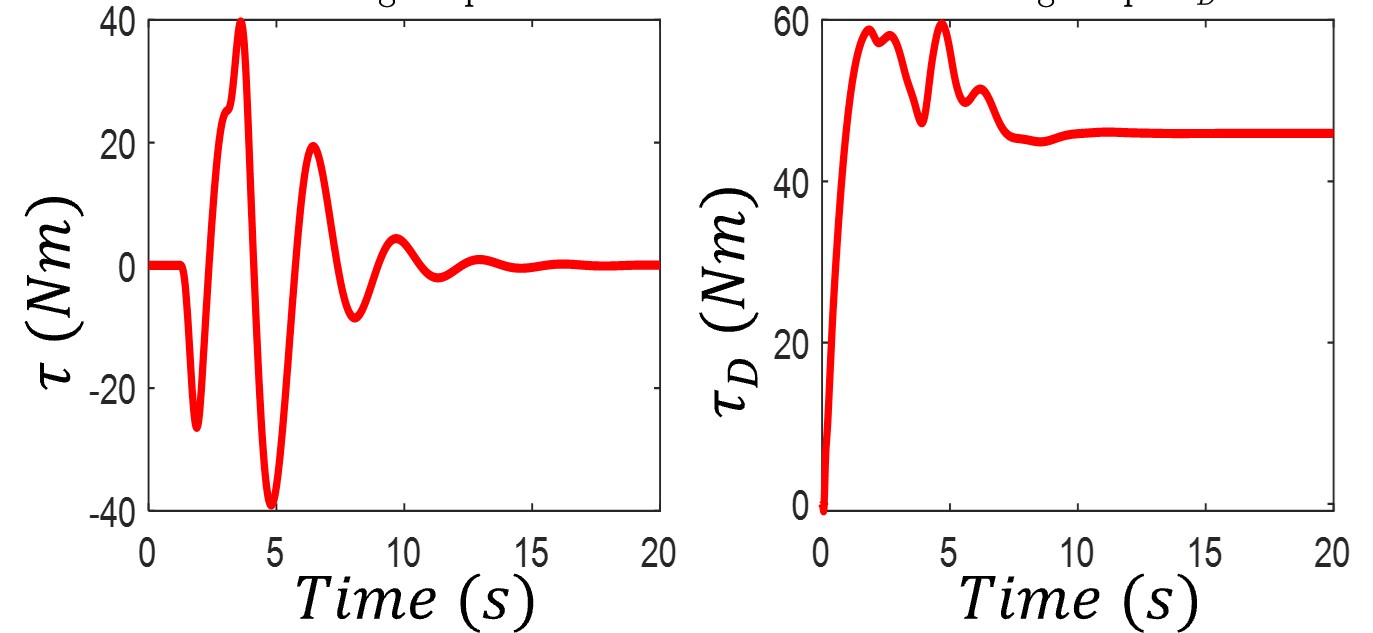}
		\caption{Input torques $\tau$, $\tau_D$}
		\label{fig12b}
	\end{subfigure}
	\hfill
	\begin{subfigure}[b]{0.235\textwidth}
		\centering
		\includegraphics[width=\textwidth]{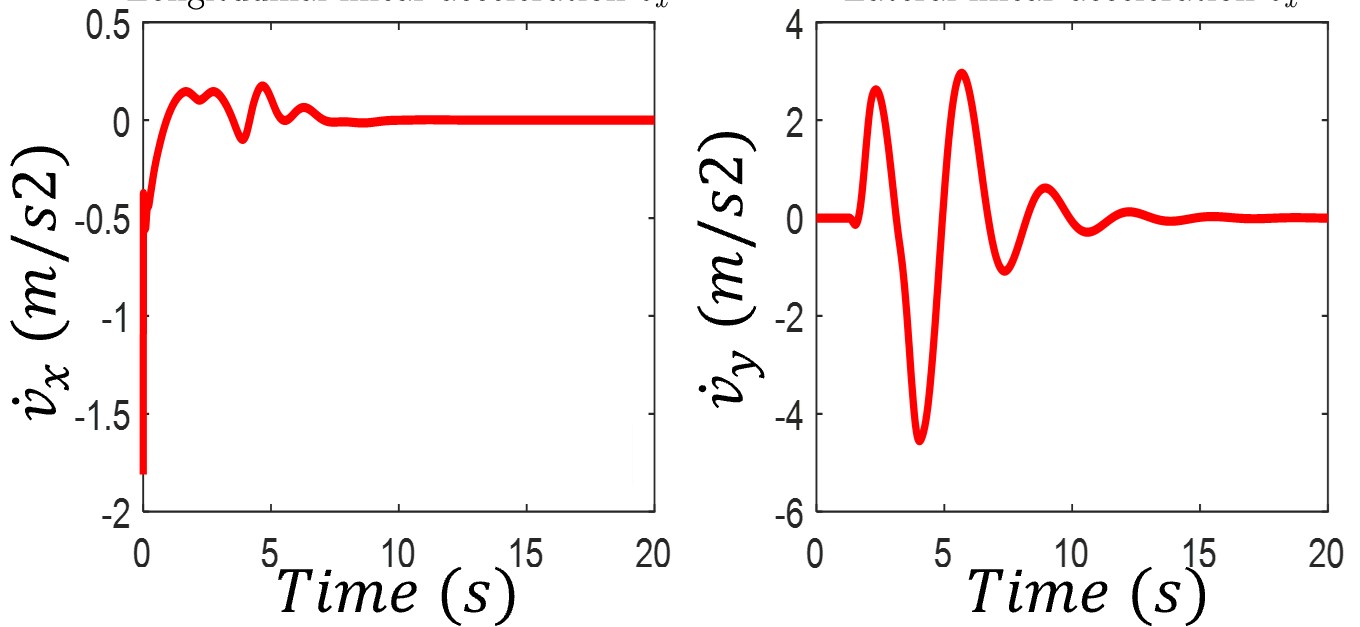}
		\caption{Linear accelerations $\dot{v}_x$, $\dot{v}_y$}
		\label{fig12c}
	\end{subfigure}
	\hfill
	\begin{subfigure}[b]{0.235\textwidth}
		\centering
		\includegraphics[width=\textwidth]{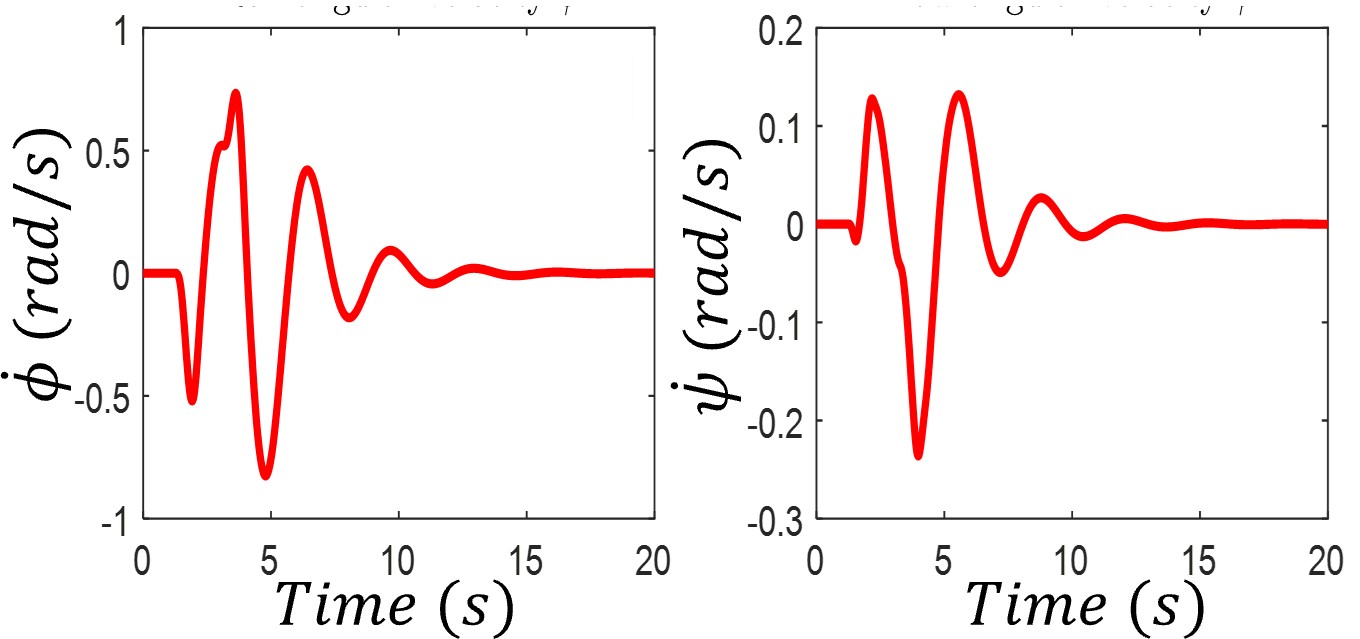}
		\caption{Angular rates $\dot{\phi}$, $\dot{\psi}$}
		\label{fig12d}
	\end{subfigure}
	\caption{Overtaking scenario at 100 kph}
	\label{fig12}
\end{figure}

For each of the scenarios, we have tested several nominal speeds. For each nominal speed and equilibrium point $(X^{*},u^{*})$, we have computed the model (\ref{Eq40c}) and designed a specific gain $G$ for the observer (\ref{Eq50a}). For each scenario, we have compared the estimation obtained on our observer with realistic simulations of BikeSim.

\subsubsection{Rectilinear trajectories.}
\begin{figure*} 
	\centering
	\begin{subfigure}[b]{0.19\textwidth}
		\centering
		\includegraphics[width=\textwidth]{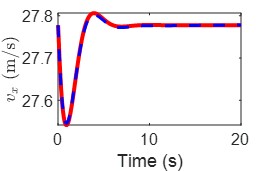}
		\caption{$v_{x}$ at 100 kph}
		\label{100a}
	\end{subfigure}
	\hfill
	\begin{subfigure}[b]{0.19\textwidth}
		\centering
		\includegraphics[width=\textwidth]{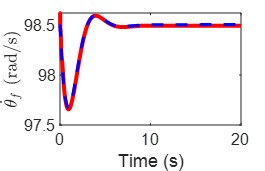}
		\caption{$\dot{\theta}_{f}$ at 100 kph}
		\label{100b}
	\end{subfigure}
	\hfill
	\begin{subfigure}[b]{0.19\textwidth}
		\centering
		\includegraphics[width=\textwidth]{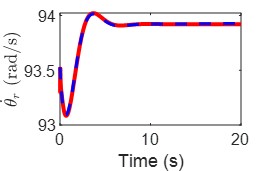}
		\caption{$\dot{\theta}_{r}$ at 100 kph}
		\label{100c}
	\end{subfigure}
	\hfill
	\begin{subfigure}[b]{0.19\textwidth}
		\centering
		\includegraphics[width=\textwidth]{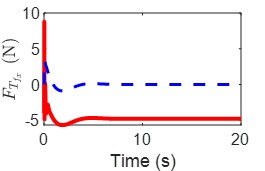}
		\caption{$F_{T_{f_x}}$ at 100 kph}
		\label{100d}
	\end{subfigure}
	\hfill
	\begin{subfigure}[b]{0.19\textwidth}
		\centering
		\includegraphics[width=\textwidth]{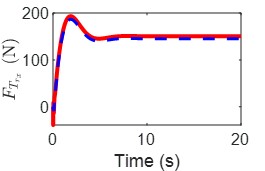}
		\caption{$F_{T_{r_x}}$ at 100 kph}
		\label{100e}
	\end{subfigure}
	\hfill
	\begin{subfigure}[b]{0.19\textwidth}
		\centering
		\includegraphics[width=\textwidth]{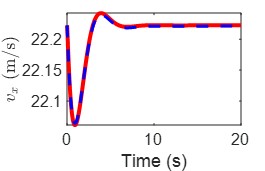}
		\caption{$v_{x}$ at 80 kph}
		\label{80a}
	\end{subfigure}
	\hfill
	\begin{subfigure}[b]{0.19\textwidth}
		\centering
		\includegraphics[width=\textwidth]{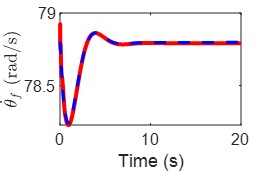}
		\caption{$\dot{\theta}_{f}$ at 80 kph}
		\label{80b}
	\end{subfigure}
	\hfill
	\begin{subfigure}[b]{0.19\textwidth}
		\centering
		\includegraphics[width=\textwidth]{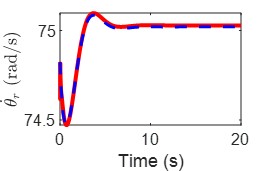}
		\caption{$\dot{\theta}_{r}$ at 80 kph}
		\label{80c}
	\end{subfigure}
	\hfill
	\begin{subfigure}[b]{0.19\textwidth}
		\centering
		\includegraphics[width=\textwidth]{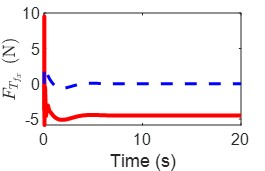}
		\caption{$F_{T_{f_x}}$ at 80 kph}
		\label{80d}
	\end{subfigure}
	\hfill
	\begin{subfigure}[b]{0.19\textwidth}
		\centering
		\includegraphics[width=\textwidth]{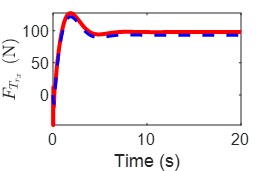}
		\caption{$F_{T_{r_x}}$ at 80 kph}
		\label{80e}
	\end{subfigure}
	\hfill
	\begin{subfigure}[b]{0.19\textwidth}
		\centering
		\includegraphics[width=\textwidth]{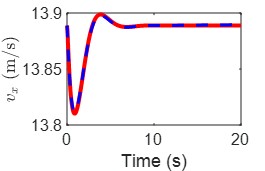}
		\caption{$v_{x}$ at 50 kph}
		\label{50a}
	\end{subfigure}
	\hfill
	\begin{subfigure}[b]{0.19\textwidth}
		\centering
		\includegraphics[width=\textwidth]{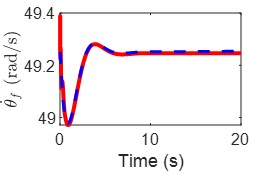}
		\label{50b}
	\end{subfigure}
	\hfill
	\begin{subfigure}[b]{0.19\textwidth}
		\centering
		\includegraphics[width=\textwidth]{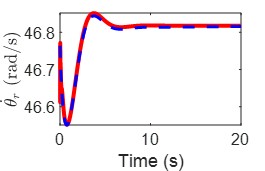}
		\label{50c}
	\end{subfigure}
	\hfill
	\begin{subfigure}[b]{0.19\textwidth}
		\centering
		\includegraphics[width=\textwidth]{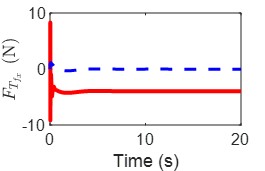}
		\label{50d}
	\end{subfigure}
	\hfill
	\begin{subfigure}[b]{0.19\textwidth}
		\centering
		\includegraphics[width=\textwidth]{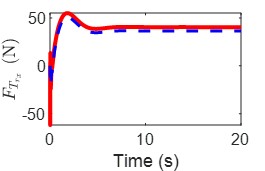}
		\label{50e}
	\end{subfigure}
	\caption{Observer dynamics for rectilinear trajectories at 100 km/h. Red lines: curves from BikeSim. Blue dashed lines: observer results.}
	\label{fig10}
\end{figure*}

We tested the observer for 100, 80 and 50 kilometers per hour (kph). For each speed, the initial value is assumed to be at the nominal point and the driving torque ($\tau_{D}$) is applied to maintain a rectilinear trajectory. The steering torque ($\tau$), front and rear braking torques ($\tau_{B_f}$ and $\tau_{B_r}$) are zero.
For each of the desired speed, the observer state is initialized as if there was no slippage ($\hat{v}_{x_0} = v_{x}^{*} = \hat{\theta}_{f_0} R_f = \hat{\theta}_{r_0} R_r$) and all the other dynamics are initialized at zero. Fig.\ref{fig10} illustrates the observer states in comparison with BikeSim output signals for 100 km/h. The results for other speeds are comparable and are presented. One can notice that the longitudinal tire forces are reconstructed, with an absolute static error less then 5 Newtons. This error is mainly due to Assumptions $A_{3}$ and $A_{4}$.

We also tested the observer designed at 80 kph for 100 and 50 kph. As it was to be expected, the performance of the observer decreases as the speed gets further from the one for which the observer was designed. Nevertheless, the observer is able to reconstruct the physical variables, such as linear velocities, angular velocities of the wheel and tire forces with a reasonable static error.  
For instance, at speed of 100 km/h (25\% deviation from the nominal speed), we observed 2\% static error on the linear velocities, while  for 50km/h (37\% deviation from the nominal speed), we observed 12\% static error. For the other nominal speeds, the experiment results are similar.
\begin{figure*}[h]
	\centering
	\begin{subfigure}[b]{0.19\textwidth}
		\centering
		\includegraphics[width=\textwidth]{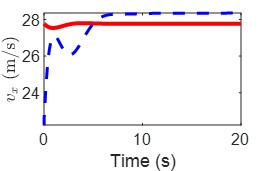}
		\caption{$v_{x}$ at 100 kph}
		\label{80-100a}
	\end{subfigure}
	\hfill
	\begin{subfigure}[b]{0.19\textwidth}
		\centering
        		\includegraphics[width=\textwidth]{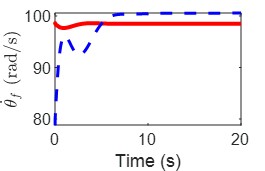}
		\caption{$\dot{\theta}_{f}$ at 100 kph}
		\label{80-100b}
	\end{subfigure}
	\hfill
	\begin{subfigure}[b]{0.19\textwidth}
		\centering
		\includegraphics[width=\textwidth]{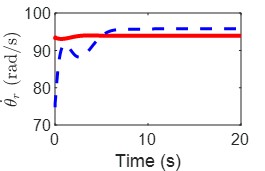}
		\caption{$\dot{\theta}_{r}$ at 100 kph}
		\label{80-100c}
	\end{subfigure}
	\hfill
	\begin{subfigure}[b]{0.19\textwidth}
		\centering
		\includegraphics[width=\textwidth]{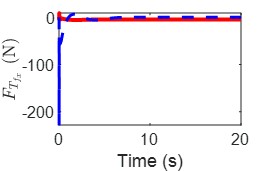}
		\caption{$F_{T_{f_x}}$ at 100 kph}
		\label{80-100d}
	\end{subfigure}
	\hfill
	\begin{subfigure}[b]{0.19\textwidth}
		\centering
		\includegraphics[width=\textwidth]{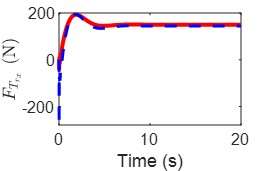}
		\caption{$F_{T_{r_x}}$ at 100 kph}
		\label{80-100e}
	\end{subfigure}
	\hfill
	\begin{subfigure}[b]{0.19\textwidth}
		\centering
		\includegraphics[width=\textwidth]{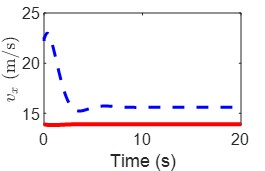}
		\caption{$v_{x}$ at 50 kph}
		\label{80-50a}
	\end{subfigure}
	\hfill
	\begin{subfigure}[b]{0.19\textwidth}
		\centering
		\includegraphics[width=\textwidth]{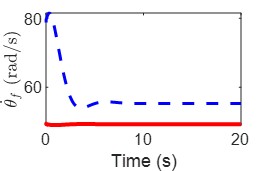}
		\caption{$\dot{\theta}_{f}$ at 50 kph}
		\label{80-50b}
	\end{subfigure}
	\hfill
	\begin{subfigure}[b]{0.19\textwidth}
		\centering
		\includegraphics[width=\textwidth]{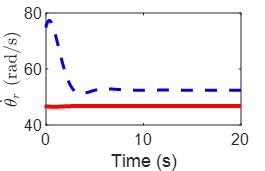}
		\caption{$\dot{\theta}_{r}$ at 50 kph}
		\label{80-50c}
	\end{subfigure}
	\hfill
	\begin{subfigure}[b]{0.19\textwidth}
		\centering
		\includegraphics[width=\textwidth]{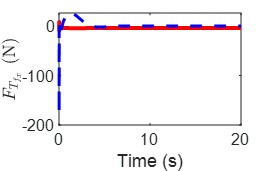}
		\caption{$F_{T_{f_x}}$ at 50 kph}
		\label{80-50d}
	\end{subfigure}
	\hfill
	\begin{subfigure}[b]{0.19\textwidth}
		\centering
		\includegraphics[width=\textwidth]{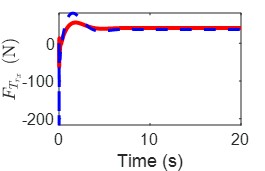}
		\caption{$F_{T_{r_x}}$ at 50 kph}
		\label{80-50e}
	\end{subfigure}
	\caption{Observer designed for 80 kph, tested at 100 and 50 kph. Red lines: curves from BikeSim. Blue dashed lines: observer results.}
	\label{fig11}
\end{figure*}

\begin{figure*}[h]
	\centering
	\begin{subfigure}[b]{0.19\textwidth}
		\centering
		\includegraphics[width=\textwidth]{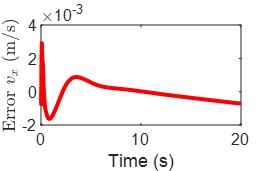}
		\caption{Error on $v_{x}$}
		\label{Error50a13}
	\end{subfigure}
	\hfill
	\begin{subfigure}[b]{0.19\textwidth}
		\centering
		\includegraphics[width=\textwidth]{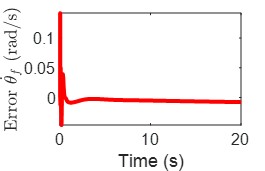}
		\caption{Error on $\dot{\theta}_{f}$}
		\label{Error50b13}
	\end{subfigure}
	\hfill
	\begin{subfigure}[b]{0.19\textwidth}
		\centering
		\includegraphics[width=\textwidth]{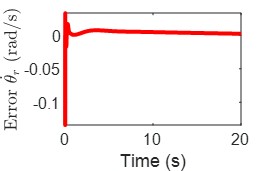}
		\caption{Error on $\dot{\theta}_{r}$}
		\label{Error50c13}
	\end{subfigure}
	\hfill
	\begin{subfigure}[b]{0.19\textwidth}
		\centering
		\includegraphics[width=\textwidth]{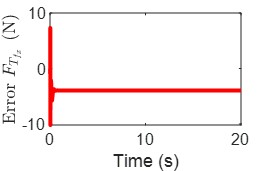}
		\caption{Error on $F_{T_{f_x}}$}
		\label{Error50d13}
	\end{subfigure}
	\hfill
	\begin{subfigure}[b]{0.19\textwidth}
		\centering
		\includegraphics[width=\textwidth]{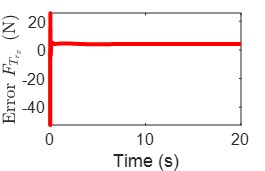}
		\caption{Error on $F_{T_{r_x}}$}
		\label{Error50e13}
	\end{subfigure}
	\caption{Observer error dynamics for 50 kph in presence of 30\% rider mass variation. Red lines: curves from BikeSim. Blue dashed lines: observer results.}
	\label{fig13}
\end{figure*}

In Fig.\ref{fig13}, we also tested our observer for robustness with respect to perturbation in model parameters. The observer error dynamics in presence of 30\% rider mass variation showed an error less than 0.1\% compared to the estimated states variables with nominal mass values.

\subsubsection{Overtaking scenario.}
Overtaking scenario induces lateral dynamics which are not taken into account in the observer design.
In this scenario, the equilibrium point ($X^{*}$,$u^{*}$) used for the observer design is the same as for the rectilinear motion at 100 kph. 
The input torques are taken from BikeSim. The braking torques are zero. The input torques generated by BikeSim ensure that at time $t$ the bike changes lanes in lateral motion.
Fig.\ref{fig73} compares the observer states with the output of BikeSim. The observer estimates all the states with a reasonable precision.

\begin{figure}
	\centering
	\begin{subfigure}[b]{0.235\textwidth}
		\centering
		\includegraphics[width=\textwidth]{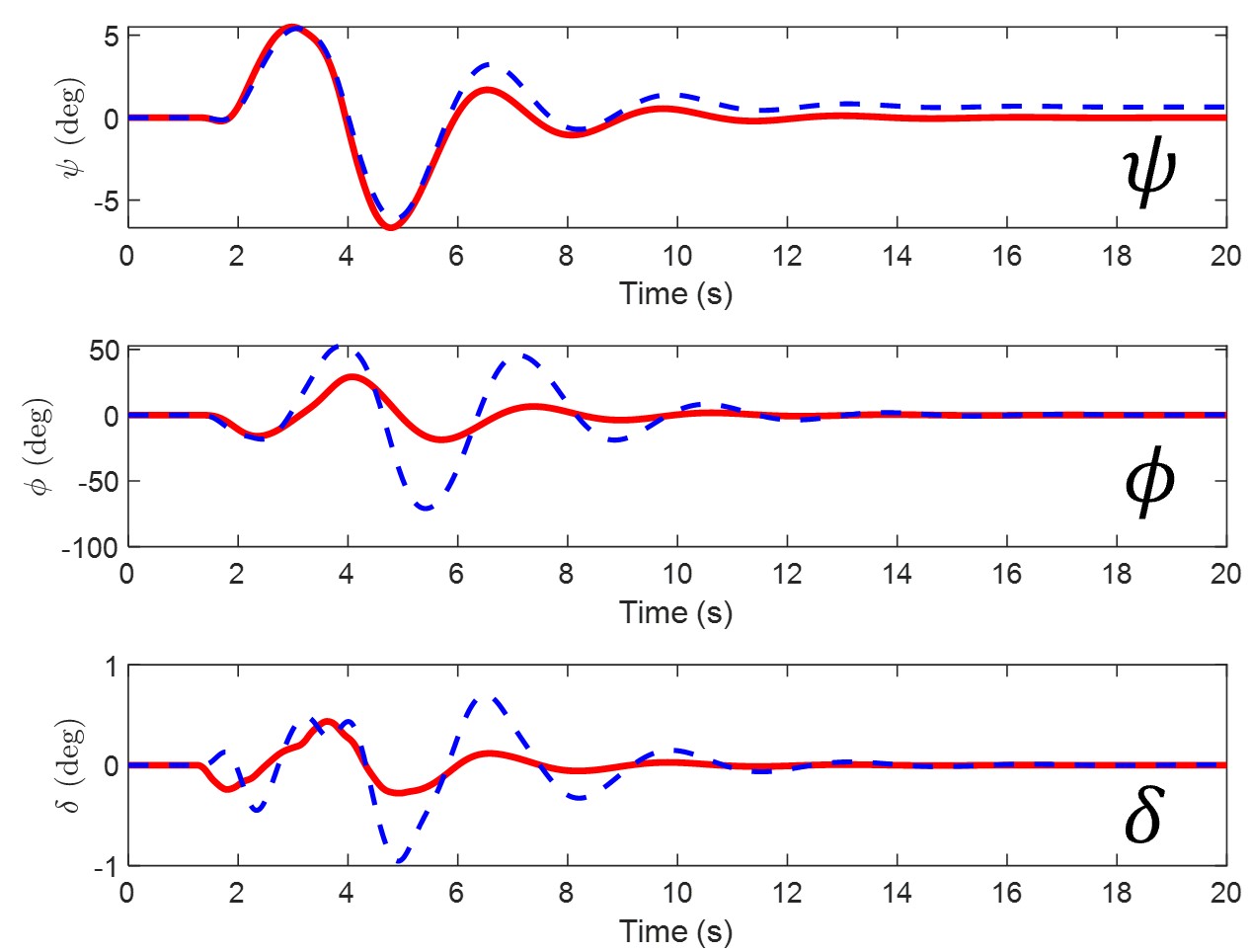}
		\caption{Lateral angles ${\psi}$, ${\phi}$, ${\delta}$.}
		\label{fig73d}
	\end{subfigure}
	\hfill
	\begin{subfigure}[b]{0.235\textwidth}
		\centering
		\includegraphics[width=\textwidth]{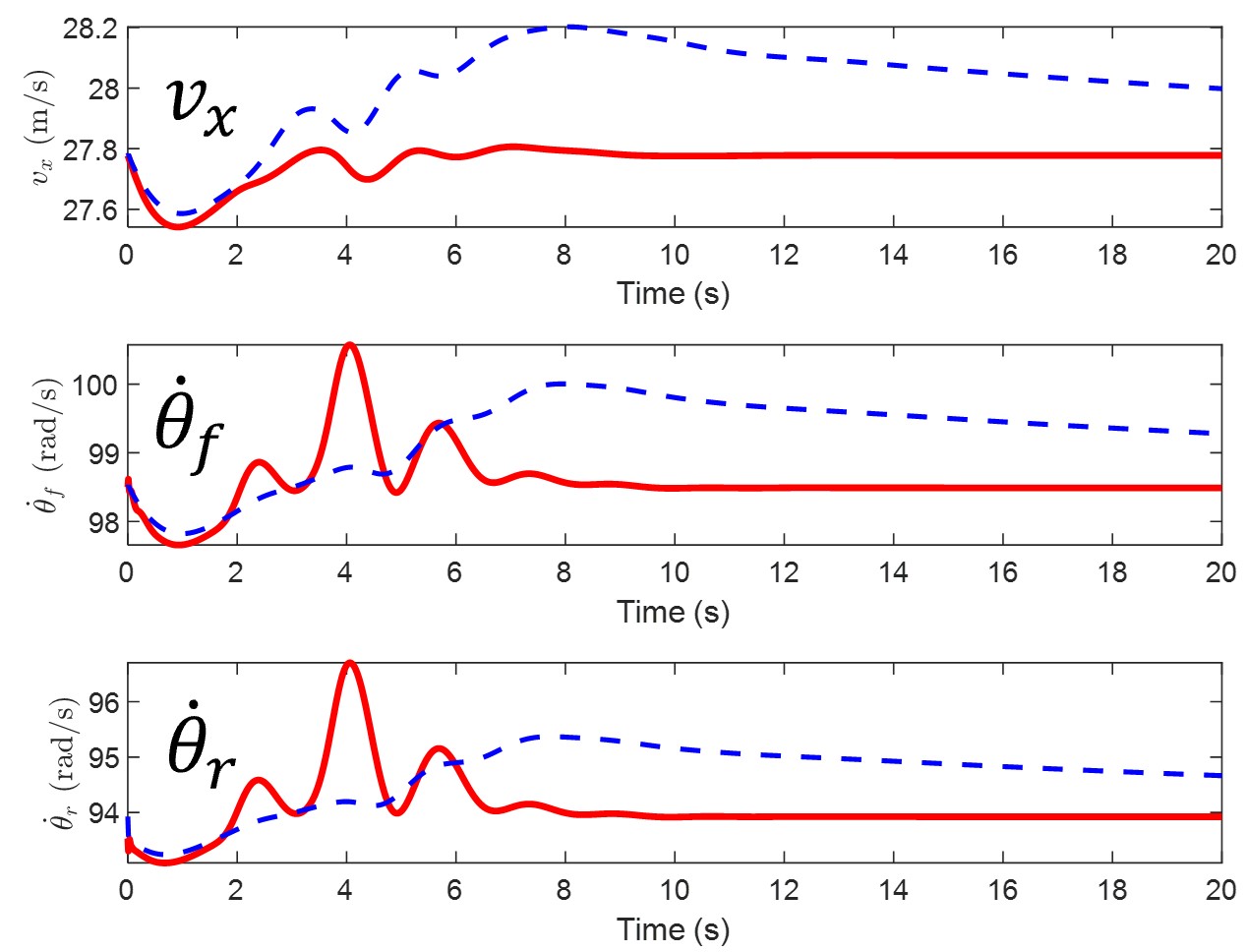}
		\caption{Longitudinal dynamics $v_{x}$, $\dot{\theta}_{f}$, $\dot{\theta}_{r}$.} 
		\label{fig73a}
	\end{subfigure}
	\caption{Observer dynamics at 100 kph overtaking scenario. Red lines: curves from BikeSim. Blue dashed lines: observer results.}
	\label{fig73v}
\end{figure}
\begin{figure}
	\begin{subfigure}[b]{0.235\textwidth}
		\centering
		\includegraphics[width=\textwidth]{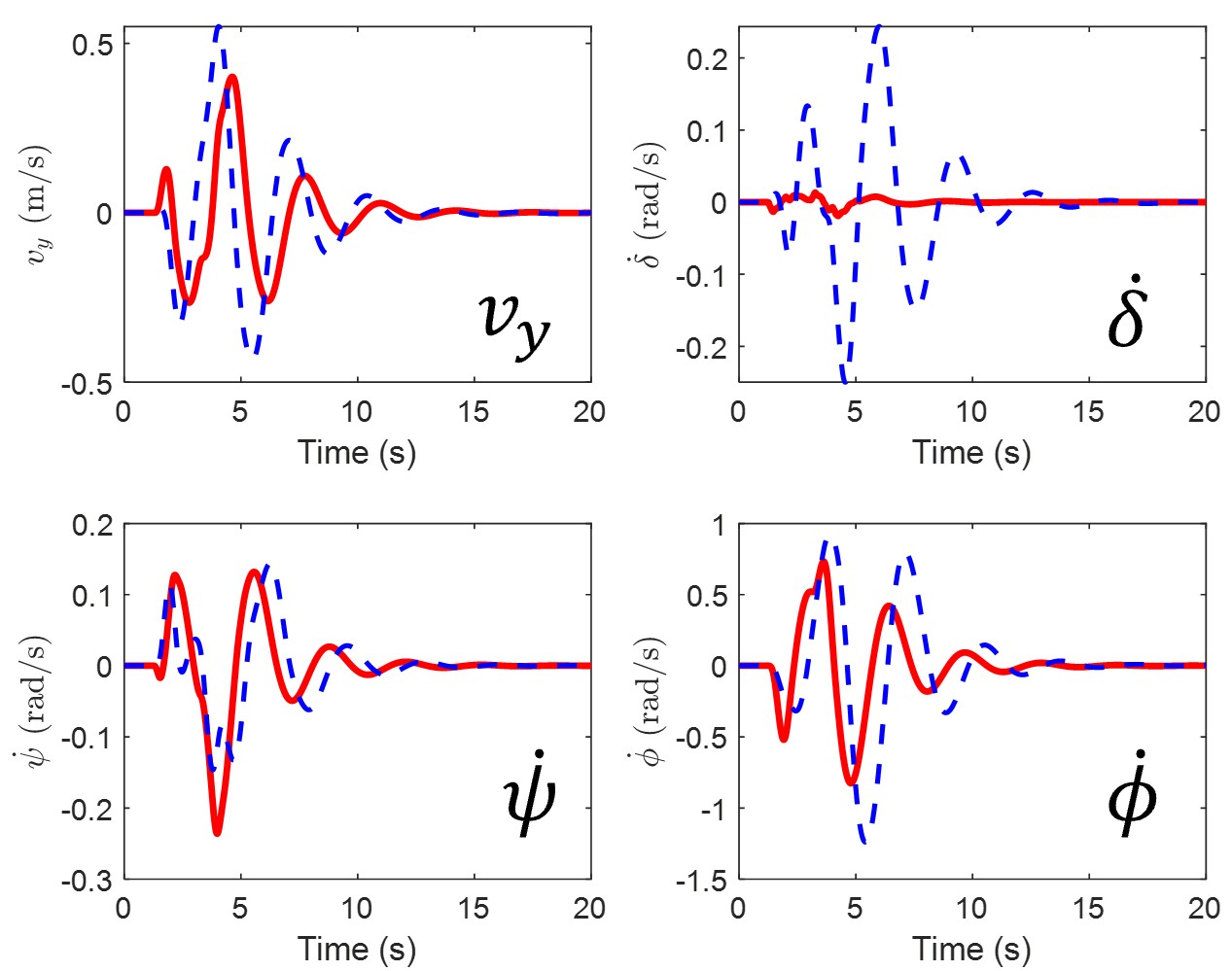}
		\caption{Lateral dynamics $v_{y}$, $\dot{\delta}$, $\dot{\psi}$, $\dot{\phi}$.}
		\label{fig73b}
	\end{subfigure}
	\hfill
	\begin{subfigure}[b]{0.235\textwidth}
		\centering
		\includegraphics[width=\textwidth]{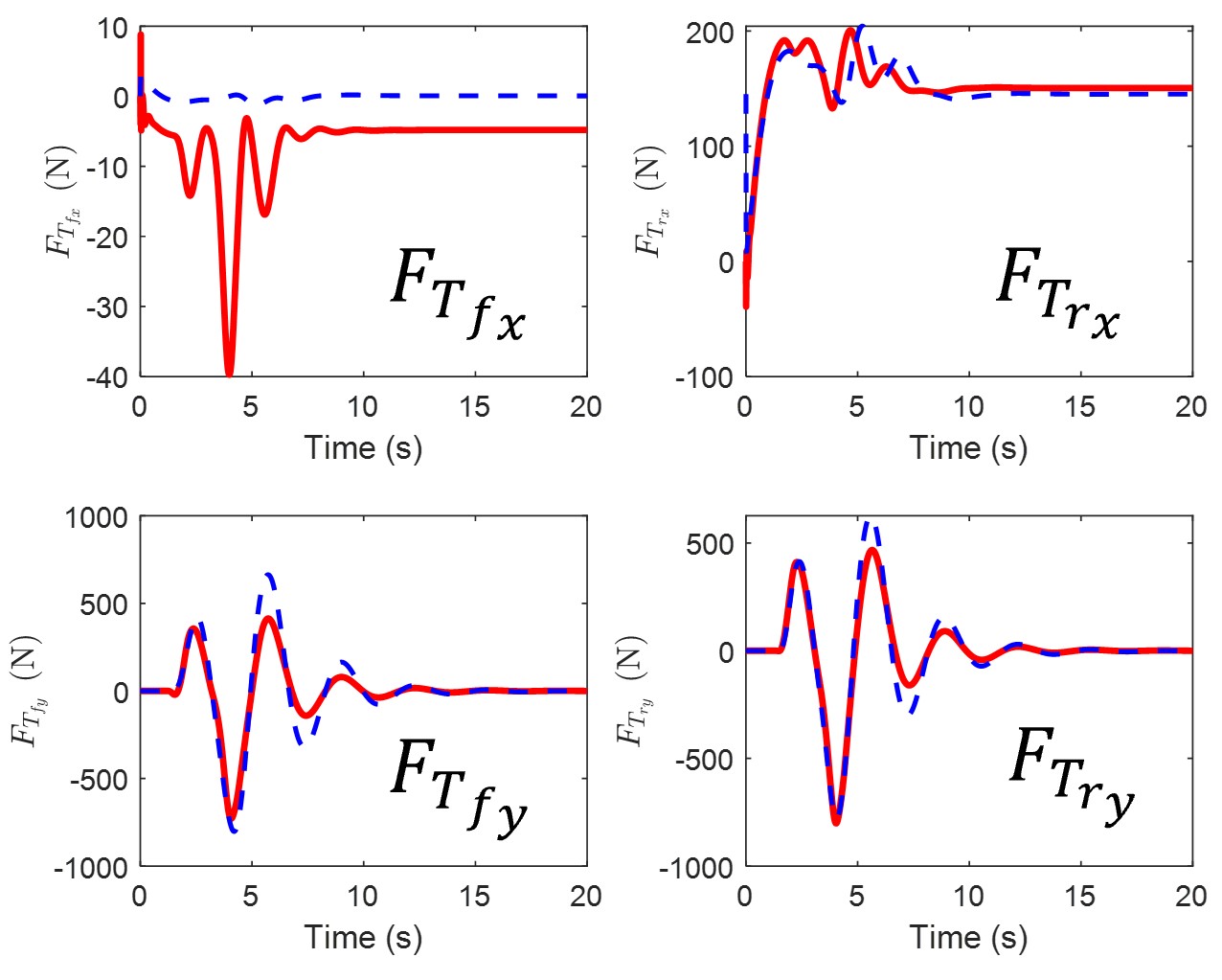}
		\caption{Tire dynamics $F_{T_{f_x}}$, $F_{T_{r_x}}$, $F_{T_{f_y}}$, $F_{T_{r_y}}$.}
		\label{fig73c}
	\end{subfigure}
	\caption{Observer dynamics at 100 kph overtaking scenario. Red lines: curves from BikeSim. Blue dashed lines: observer results.}
	\label{fig73}
\end{figure}

\section{Conclusion}
This paper introduces a four-bodies dynamic model for motorcycles based of Jourdain's principle, considering both longitudinal and lateral dynamics. The analytical model presented was used to design a full states observer based on linear quadratic regulator theory. The results were compared with BikeSim simulator outputs. This highlights the potential of the presented model to improve motorcycle safety measures in various driving scenarios where only inertial measurement unit sensors are available.

Further works involving an enhanced control oriented model that additionally gathers vertical variables (such as pitch and suspension deflections) are under investigation. This could potentially lead to estimating PTW dynamics during more types of maneuvers.

\begin{ack}
This work is part of an industrial project funded by Autoliv and Agence Nationale de la Recherche et de la Technologie (ANRT). Autoliv Electronics and Autoliv Research are part of Autoliv (www.autoliv.com), the worldwide leader in automotive safety systems.
\end{ack}

\bibliography{bibliography}             

\end{document}